# Non-Isothermal Model for Nematic Spherulite Growth

*Nasser Mohieddin Abukhdeir,[a]\* Ezequiel R. Soulé,[a,b] and Alejandro D. Rey[a]*

[a]: Department of Chemical Engineering, McGill University, Montreal, Quebec H3A 2B2

[b]: Institute of Materials Science and Technology (INTEMA), University of Mar del Plata and National Research Council (CONICET), J. B. Justo 4302, 7600 Mar del Plata, Argentina

*To whom correspondence should be addressed. E-mail: nasser.abukhdeir@mcgill.ca

**ABSTRACT**

A computational study of the growth of two-dimensional nematic spherulites in an isotropic phase was performed using a Landau-de Gennes type quadrupolar tensor order parameter model for the first-order isotropic/nematic transition of 5CB (pentyl-cyanobiphenyl). An energy balance, taking anisotropy into account, was derived and incorporated into the time-dependent model. Growth laws were determined for two different spherulite morphologies of the form $t^n$, with and without the inclusion of thermal effects. Results show that incorporation of the thermal energy balance correctly predicts the transition of the growth law exponent from the volume driven regime ($n$=1) to the thermally limited regime (approaching $n=\frac{1}{2}$), agreeing well with experimental observations. An interfacial nemato-dynamic model is used to gain insight into the interactions that result in the progression of different spherulite growth regimes.

**KEYWORDS**: liquid crystal, nematic, phase transition, growth law, latent heat

**INTRODUCTION**

The kinetics of phase transformations[1] is a fundamental subject in material and interfacial science that has widespread impact in material manufacturing and use. Three important kinetic phenomena associated with phase transformations are microstructure, growth rate, and shape evolution.[2] Phase transformations usually lead to microstructures, such as dendrites in metals[3-5] and anisotropic spherulites in semi-crystalline polymers[6] among other examples.[7] The growth processes associated with phase transformations vary, depending on driving forces, such as bulk free energy minimization and reduction of interfacial area. Phase transformations are usually classified into diffusive and non-diffusive, according to whether heat and/or mass transport are involved.[1,2] Growth laws in diffusive and non-diffusive transformations typically differ in that the former has a conserved order parameter.[1,2] Phase ordering in nematic liquid crystalline materials[8] is an important phase transformation giving rise to microstructures such as defects in spherulites as well as complex growth laws.[9-12] When analyzing phase ordering in nematic liquid crystals, latent heat is typically neglected and, subsequently, the transformation is considered to be diffusionless.[9-12] Recently, evidence has shown that thermal effects are significant and heat diffusion must be taken into account to reproduce the observed growth laws.[13] Thus the minimal description of isotropic/nematic phase transformations via spherulite growth requires the description of four coupled processes: shape kinetics, growth kinetics, texturing, and heat transport.

• Shape kinetics: the shape kinetics of a growing spherulite is specified by the transport law:[14-16]

$$\frac{d\mathbf{b}}{dt} = \nabla_s \left[ (\nabla_s \mathbf{v} \cdot \mathbf{k}) \right] - (\nabla_s \mathbf{v}) \cdot \mathbf{b} + \mathbf{k}\mathbf{b} \cdot (\nabla_s \mathbf{v}) \cdot \mathbf{k} \qquad (1)$$

where $\mathbf{v}$ is the interface velocity, $\mathbf{b} = -\nabla_s \mathbf{k}$ is the curvature tensor, $\mathbf{k}$ is the unit surface normal, $\nabla_s = (\mathbf{I} - \mathbf{k}\mathbf{k}) \cdot \nabla$ is the surface gradient tensor, $\mathbf{I}$ is three-dimensional unit dyadic, and $\nabla$ is the three-dimensional gradient operator. For two-dimensional spherulite growth where the normal interface velocity is $\mathbf{v} = w\mathbf{k}$, eq 1 simplifies to:

$$\frac{dH}{dt} = 2H^2 w + \frac{1}{2} \nabla_s^2 w \qquad (2)$$

where H is the average curvature.[14] Equation 2 shows that under constant normal velocity and two-dimensional growth, the radius of a spherulite R (where H=-1/R) follows a linear law: $R \propto 2wt$. When the normal velocity is not constant, the radius obeys:

$$\frac{dR}{dt} - 2H^2 w - \frac{R^2}{2}\nabla^2_s w = 0 \qquad (3)$$

and thus R is not a linear function of time.

• Growth kinetics: the normal velocity of the interface w is usually dictated by:[2,10-12]

$$\beta w = \Delta F(T_s - T^*) - \mathbb{C} \qquad (4)$$

where β is the interfacial viscosity, $\Delta F(T_s - T^*)$ is the driving force (free energy difference), $T_s$ is the interfacial temperature, $T^*$ is the transition temperature, $T_s - T^*$ is under-cooling, and $\mathbb{C}$ is the resisting capillary force. For diffusionless transformations $T_s$ is the constant quench temperature, but for non-negligible heat of transitions $T_s$ changes with time and hence w is not constant.

• Texturing: during phase ordering, heterogeneities in orientational and molecular order arise within the nematic spherulite, including defects.[9-12] Since the dynamics of the liquid crystalline order[3] are dictated by the variation of the temperature-dependent free energy, latent heat again has an effect.

• Heat transfer; since the latent heat evolved during the isotropic/nematic transition is not insignificant,[13] the temperature of the spherulite and its interface increases.

Based on the above discussion, the minimum model to realistically describe mesophase growth must take into account the processes and couplings shown in Figure 1.

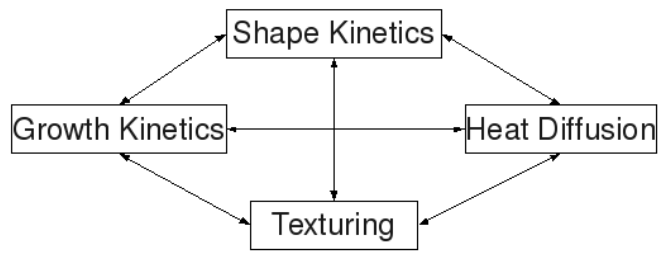

**Figure 1**: Schematic showing the inter-relationships between the phase ordering processes.

By including heat transfer processes into the isotropic/nematic phase transformation, a generalized phase ordering model is realized. Figure 2 shows a schematic of different types of phase ordering models. The left stream is the isothermal model that neglects latent heat and takes into account growth (**v**), shape (**b**), and textures (**Q**); these models have been widely studied in the past but only capture some of the experimentally observed growth regimes. The right stream is the non-isothermal model that considers the effect of heat diffusion ($\nabla T$) on liquid crystalline phase order texturing (**Q**), growth (**v**), and shape kinetics (**b**). The (**Q**,$T$) couplings in the non-isothermal model have not been fully characterized and the need to increase this understanding motivates this paper.

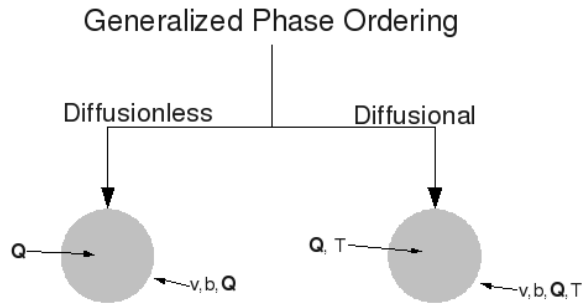

**Figure 2**: Classification of phase ordering processes: diffusionless (isothermal) and diffusional (non-isothermal). **Q** is the tensor order parameter, **v** is growth velocity, **b** is the curvature tensor, and is T temperature.

Experimental studies on growth of liquid crystalline spherulites are not abundant.[17-22] This past work[17-22] on the growth of liquid crystalline spherulites sheds light on macroscopic growth laws and, to some extent, morphology. The underlying mechanisms and mesoscopic dynamics are difficult to access experimentally and computer simulation is a useful complementary tool to experimental studies.

Recently, Huisman and Fasolino[13] showed that commonly neglected thermal effects (due to the relatively low latent heat of these types of transformations, when compared to solidification, for example), are indeed relevant to the growth of nematic spherulites. Like many natural growth processes, the growth of a nematic spherulite in an isotropic matrix exhibits self-similarity, where the spherulite radius obeys a power law relationship with time, $t^n$. When the quench conditions are "deep," meaning that the quench temperature is far below the lower stability limit of the isotropic phase, $n=1$ once spherulite growth enters the self-similar regime. This type of growth is driven by the volumetric change in the free energy due to the phase transition and is referred to as volume-driven growth. As the quench temperature is raised, the quench become more "shallow" and the growth exponent is experimentally observed to approach $n=\frac{1}{2}$. This change has been attributed to the decrease in magnitude of the free energy drop of the phase transition and growth proceeding through capillary forces, referred to as curvature growth.[13] Huisman and Fasolino[13] argue that this view is incorrect; curvature effects can only inhibit interface growth and that the change in exponent is a result of latent heat effects. They argue that the latent heat released by the moving phase-transition front increases the local temperature, slowing growth. For deep quenches, temperature increases at the interface have a negligible effect on the free energy difference, allowing the interface to grow at a constant velocity, as found by the diffusionless model.[9-12] These changes are magnified for shallow quenches in that latent heat effects are able to increase the interface temperature such that it approaches the bulk transition temperature, thus limiting growth by thermal diffusion. They show[13] with a simple scalar model and the assumption that the interface grows in a shape-preserving way, that the transition for power law exponent $n=1$ to $n=\frac{1}{2}$ is found, if and only, thermal effects are taken into account.

In this work we extend the tensorial Landau-de Gennes model for the isotropic/nematic phase transition by the addition of an anisotropic thermal energy balance. We unequivocally demonstrate that by incorporating latent heat effects into this model, the isothermal Landau-de Gennes predictions[9-12] are modified and the growth law exponent $n$ is found to transition from $n=1$ to approaching $n=\frac{1}{2}$, as observed experimentally. The non-isothermal simulation approach allows for a more complete spectrum of growth phenomena to be captured. This includes prediction of the transition mechanism from early spherulite growth to steady-state interfacial growth with details inaccessible experimentally due to the length scales (nanometers) and time scales (nanoseconds) involved. The single spherulite non-isothermal growth laws that we seek to establish will then give the necessary foundations to future studies on three-dimensional growth in multi-spherulite systems. It should be noted that even using parallel adaptive-mesh finite element solution techniques, currently available computational power is not adequate to resolve statistically significant multiple spherulite ensembles produced in experiments. The main emphases of this paper are growth laws and heat transfer, while shape kinetics and texturing processes are only included for completeness and will be treated in detail in future work.

The objectives of this work are 1) to provide a necessary bridge from the simplified scalar model used by Huisman and Fasolino[13] to a more fine-grained three-dimension model and 2) to show that the incorporation of latent heat effects corrects the standard prediction of the tensorial Landau-de Gennes model from $n=1$ towards $n=\frac{1}{2}$ for low molar mass calamitic thermotropic liquid crystals.

Objective 1 is accomplished using a standard approach of liquid crystal physics to extend an isothermal tensorial Landau-de Gennes model for the first-order isotropic/nematic phase transition.[8,23-27] This type of mesoscopic model has the benefit of being relatively simple, yet capturing the core phenomena involved in the isotropic/nematic phase transition.[24,28] Other modeling approaches, including molecular dynamics simulations, are in use to simulate this transition[29-33] but are currently not feasible for accessing the length scales (microns) that are of interest in this work. The use of a full tensorial model captures the required phenomena for realistic simulation of the isotropic/nematic transition (see Figure 1). As will be shown, the presented model fully incorporates the nontrivial texture/shape kinetics involved in the growth of nematic spherulites[9-12] and verifies influence of thermal diffusion in the presence of the full spectrum of simultaneously occurring spherulite growth processes (see Figure 1).

Objective 2 is accomplished by simulation in two-dimensions of the presented model. Specifically, this work seeks to confirm the experimental[17-22] and theoretical[13] results using a more complete non-isothermal model than that used by Huisman and Fasolino.[13]

The paper is organized as follows: the Landau-de Gennes type free energy and anisotropic thermal energy balance are introduced, the simulation conditions and assumptions are explained, and simulation results presented and discussed.

**BACKGROUND AND SIMULATION APPROACH**

The Landau-de Gennes quadrupolar tensor order parameter model for the first-order isotropic/nematic phase transition is a polynomial expansion of the free energy change due to nematic phase ordering. This order is characterized by the symmetric-traceless quadrupolar tensor:[8]

$$\mathbf{Q} = \mu_n \mathbf{nn} + \mu_m \mathbf{mm} + \mu_l \mathbf{ll} \tag{5a}$$

$$\mu_n + \mu_m + \mu_l = 0 \tag{5b}$$

where orientational order is characterized by **n**, **m**, and **l**, the tensor eigenvectors, and the extent to which the molecules conform to this order by $\mu_n$, $\mu_m$, and $\mu_l$, the tensor eigenvalues. Thus, isotropic order corresponds to all eigenvalues equaling zero. When uniaxial nematic order is present, all of the eigenvalues are nonzero with two being equal. The third eigenvalue is referred to as the uniaxial scalar nematic order parameter, S, and the eigenvector associated with it the nematic director, **n**. In the case of biaxial nematic order, all eigenvalues are nonzero and differ (see ref. 9 for more details). The free-energy density is composed of bulk and elastic contributions:

$$f - f_i = f_n = f_h(\mathbf{Q}) + f_g(\mathbf{Q}, \nabla \mathbf{Q}) \tag{6a}$$

$$F = \int_V (f_n) dV \tag{6b}$$

where $f$ is the total free energy density, $f_i$ is the free energy density of the isotropic phase, $f_h$ is the homogeneous contribution of nematic ordering, $f_g$ is the elastic contribution, and $F$ is the total free energy. The Landau-type free energy expansion includes:[9]

$$f_n = f_h + f_g \tag{7a}$$

$$f_h = \frac{1}{2} a (\mathbf{Q}{:}\mathbf{Q}) - \frac{1}{3} b (\mathbf{Q}{\cdot}\mathbf{Q}){:}\mathbf{Q} + \frac{1}{4} c (\mathbf{Q}{:}\mathbf{Q})^2 ; f_g = \frac{1}{2} l_1 (\nabla \mathbf{Q} {\vdots} \nabla \mathbf{Q}) + \frac{1}{2} l_2 (\nabla {\cdot} \mathbf{Q}) {\cdot} (\nabla {\cdot} \mathbf{Q}) + l_3 \mathbf{Q}{:}(\nabla \mathbf{Q}{:}\nabla \mathbf{Q}) \tag{7b}$$

$$a = a_0 (T - T_{NI}) \tag{7c}$$

where $a$, $b$, $c$ are bulk parameters, $T_{NI}$ is the lower stability limit of the isotropic phase, and $l_1$, $l_2$, $l_3$ are the tensorial elastics constants. An equi-bend/splay assumption is used in this work, resulting in $l_1$, $l_2 > 0$ and $l_3 = 0$. This assumption is made based upon previous studies of nematic spherulite morphology resulting from the interplay of splay, twist, and bend elastic constants.[9-12] Interfacial contributions implicitly result from the inclusion of the gradients terms $f_g$ in eq 7b. Detailed past work has studied the interfacial contributions of these gradient terms including anchoring, curvature, and heterogeneous effects. Refer to refs 9-12 for a comprehensive study of these effects.

The Landau-Ginzburg time dependent formulation is used to minimize the free energy functional eq 6b of the simulation volume:

$$\mu \frac{\partial \mathbf{Q}}{\partial t} = -\left[\frac{\delta F}{\delta \mathbf{Q}}\right]^{ST} \tag{8}$$

where $\mu$ is the rotational viscosity and only the symmetric-traceless component of the functional derivative is utilized (denoted by the superscript "ST").

The general differential energy balance, neglecting convection, is (derived in Appendix 1):

$$C_p \frac{\partial T}{\partial t} = \left(-\frac{\partial f}{\partial \mathbf{Q}}\bigg|_{T,\nabla \mathbf{Q}} + \nabla \cdot \frac{\partial f}{\partial \nabla \mathbf{Q}}\right) : \frac{\partial \mathbf{Q}}{\partial t} + T \frac{\partial}{\partial \mathbf{Q}} \left(\frac{\partial f}{\partial T}\bigg|_{\mathbf{Q},\nabla \mathbf{Q}}\right) : \frac{\partial \mathbf{Q}}{\partial t} - \nabla \cdot \mathbf{q} \tag{9}$$

where $C_p$ is the heat capacity, $f$ is the free energy, and $\mathbf{q}$ is the total heat flux. Temperature fluctuations are neglected in this model but could be incorporated via stochastic terms.

The heat flux can be calculated from the anisotropic Fourier's law:

$$\mathbf{q} = -\mathbf{K} \cdot \nabla T \tag{10}$$

where the thermal conductivity tensor $\mathbf{K}$ is used due to the anisotropy of the nematic phase. The thermal conductivity tensor can be written as the sum of an isotropic and anisotropic contributions:

$$\mathbf{K} = k_{iso}\mathbf{I} + k_{an}\mathbf{Q} = \left(\frac{k_{//} + 2k_{\perp}}{3}\right)\mathbf{I} + (k_{//} - k_{\perp})\mathbf{Q} \tag{11}$$

where $k_{iso}$ and $k_{an}$ are the isotropic and anisotropic contributions to the thermal conductivities, and $k_{//}$ and $k_{\perp}$ are the conductivities in the directions parallel and perpendicular to the director, respectively.

The resulting thermal energy balance eq 9 can be rewritten using eq 8 as:

$$\rho C p \left(\frac{\partial T}{\partial t}\right) = \mu \frac{\partial \mathbf{Q}}{\partial t} : \frac{\partial \mathbf{Q}}{\partial t} + T \left\{\frac{\partial}{\partial \mathbf{Q}}\left(\frac{\partial f_n}{\partial T}\right)\right\} : \frac{\partial \mathbf{Q}}{\partial t} - \nabla \cdot \mathbf{q} \tag{12}$$

where the constitutive law for the heat flux is defined in eqns 10-11. The first term on the right-hand side corresponds to dissipation, the second is the latent heat contribution of the phase transition, and the third the net heat flux (see Appendix 1 for more details).

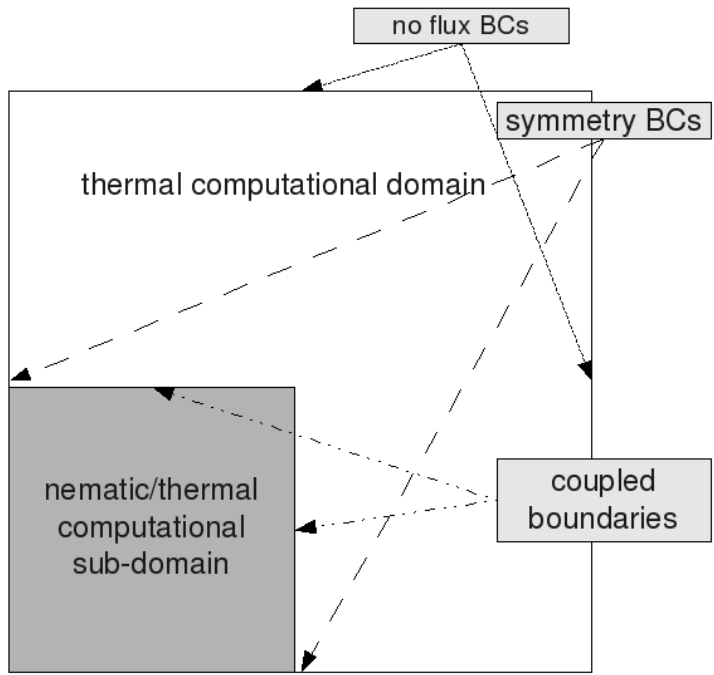

**Figure 3** – A schematic of the computational domain where nematic order and the thermal energy balance are both computed over the shaded sub-domain and only the thermal energy balance is computed over the full domain.

In this work, we use experimentally reported material property data for 5CB (pentyl-cyanobiphenyl) which will be presented in the following section.[9,10,34] The time-dependent closed model defined by eqns 8 and 12 were simulated using these material parameters and a commerical finite element package, Comsol Multiphysics.

Simulation conditions for the shallow temperature quench were guided by the phase diagram computed from the free energy density eq 7b. The geometry of the simulation domain and boundary condition types are shown in Figure 3. An inner sub-domain, with a very refined mesh, is used in order to resolve the details at the interface of the droplet. This sub-domain is enclosed by a larger domain, with a coarser mesh, where only the energy balance is solved for. This is necessary in that the length in which temperature gradients develop can be much larger than the droplet radius and using a fine mesh (which is not necessary to resolve the temperature profiles) in the full domain is not possible due to computational limitations.

The symmetry of the single 2D spherulite results in the following boundary conditions:

$$\frac{\partial Q_{xx}}{\partial x} = 0; \quad \frac{\partial Q_{yy}}{\partial y} = 0; \quad Q_{xy} = Q_{yx} = 0; \quad \frac{\partial T}{\partial y} = \frac{\partial T}{\partial x} = 0 \tag{13}$$

The off-diagonal components of the Q-tensor eq 5a are zero at the symmetry axis, which can be explained geometrically. Considering the value of one of the Q-tensor eigenvectors (**n**, **m**, or **l** from eqn 5b) that crosses an axis of symmetry in a two-dimensional field, one of the vector components must change sign across the axis and thus equal zero at the axis intersection. The same argument applies for its corresponding orthogonal eigenvector in the plane, and thus all off-diagonal Q-tensor components equal zero at this intersection in that one of the eigenvector components that composes it equals zero.

Simulations are run such that nematic spherulite growth does not exceed the sub-domain boundary. A finite element method is used to solve the coupled set of partial-differential equations where the mesh size is determined by the characteristic length scale of the nematic phase:

$$\lambda_N = \sqrt{\frac{l_1}{a}} \tag{14}$$

The Heaviside step function is used to generate the initial circular spherulite:

$$\mathbf{Q} = \mathbf{Q}^0 H(r - r_0) \tag{15}$$

where $\mathbf{Q}^0$ is the **Q**-tensor value at the bulk conditions corresponding to the quench depth. $H$ is the Heaviside step function, $r$ is the radius from the origin, and $r_0$ is the initial spherulite radius corresponding to the quench depth. Additionally, the size of the thermal domain (refer to Figure 3) was determined such that no heat flux reached the boundaries of the domain. This condition was verified post-simulation by confirming that the temperature at these boundaries remained unchanged (equaling that of the original quench temperature).

Initial spherulite nucleus conditions were divided into two cases: homogenous and textured (see Figure 4). For the initially textured case we choose the well-known radial planar texture described by a purely radial director field with homeotropic boundary conditions and a disclination line at the center of strength +1.[35-37] Isotropic conditions are used for the initial state of the spherulite center. After time $t=0$ the interfacial conditions are self-selected and not imposed, allowing for the formation of a +1 disclination at the spherulite center. Likewise, the +1 disclination line at the center of the spherulite can, in principle, remain as-is, split into two +1/2 disclination lines, or escape through the interface (in full three-dimensional simulations); hence the texturing is self-selected as indicated by the couplings in Figure 2.

Volumetric growth is present at quench depths below the lower stability limit of the isotropic phase, or "deep" quenches.[9-12] As the quench depth exceeds the isotropic stability limit, capillary effects become relevant and the initial spherulite must have a radius larger than a critical value in order to overcome the surface energy and grow. For each spherulite configuration, the critical radius for the shallow quench was determined, within an accuracy of $\pm\lambda_N$, by rerunning simulations without the thermal energy balance at increasing radii until spherulite growth proceeded. Four types of simulations were then performed: (i) isothermal model with initially homogeneous spherulite (IHS), (ii) non-isothermal model with initially homogeneous spherulite (NIHS), (iii) isothermal model with initially textured spherulite (ITS), and (iv) non-isothermal model with initially textured spherulite (NITS).

The growth rate of the spherulite is characterized by a power law of the type $R \approx t^n$ where n is the power law exponent, and for the four cases investigated, we compute $n(t)$. When the spherulite is non-circular, the long axis is used.

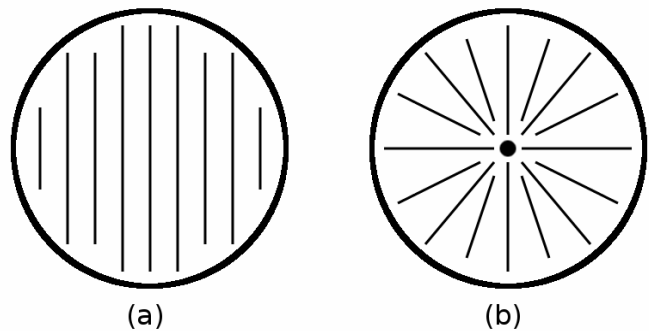

(a)      (b)

**Figure 4** – Schematics of the a) homogeneous and b) textured spherulite director orientation profiles used as initial conditions. The dark dot denotes a +1 disclination line defect. These two initial conditions were used in the isothermal and non-isothermal models.

## RESULTS AND DISCUSSION

In order to characterize the complex interfacial dynamics of the growth of a nematic spherulite of the homogeneous type, a previously derived [9,10] interfacial nemato-dynamic equation is introduced:

$$\beta w = (\mathbf{L} + \nabla_s \cdot \mathbf{T}_s) \cdot \mathbf{k} + \mu \mathbf{Q}^s : \frac{d\mathbf{Q}^s}{dt} \quad (16)$$

where $\beta$ is the surface viscosity, $w$ is the normal velocity of the spherulite interface, $\mathbf{L}$ is the bulk elastic load acting on the interface, $\nabla_s$ is the surface gradient, $\mathbf{T}_s$ is the surface stress tensor, $\mathbf{Q}^s$ is the order parameter at the interface, and $\mu$ is the rotational viscosity. Using this model, the growth regimes of a nematic spherulite can be characterized in the context of the initial nucleus configuration. Unlike that of the textured-type spherulite, homogeneous spherulite growth is dominated by a relatively prolonged shape dynamics regime, where capillary forces and transient viscous effects prevail. Following this regime, the spherulite grows in a self-similar shape which results in the spherulite radius obeying a power law relationship in time. The driving force for spherulite growth is the bulk elastic load, $\mathbf{L}$. This load is related to the free energy difference between the isotropic/nematic phases. As mentioned above, this type of growth is referred to as volumetric growth, and for a diffuse interface model[9,10] is given by:

$$\mathbf{L} \cdot \mathbf{k} = [f_h \mathrm{h}]_{\lambda=\Delta_I} - [f_h \mathrm{h}]_{\lambda=\Delta_N} \quad (17a)$$

$$\mathrm{h} = 1 - 2\mathrm{H}\lambda \quad (17b)$$

where h is a measure of the perimeter change as the diffusive interface is traversed along its normal direction and it is given by $\mathrm{h} = 1 - 2\mathrm{H}\lambda$ in a two-dimensional geometry, $\lambda$ is the coordinate measured from the center of the interface along the unit normal, and $\lambda = \Delta_I, \lambda = \Delta_N$ are the locations of the isotropic and nematic bulk phases, respectively. The interface thickness corresponds to $|\Delta_I - \Delta_N|$. Note that for a flat surface H=0 and the area amplification factor h (see eq 17b) is h=1. In essence, the load $\mathbf{L \cdot k}$ is the temperature-dependent free energy difference between the bulk isotropic phase and the nematic phase evaluated at the interface temperature:

$$\mathbf{L \cdot k} = -\left(\frac{1}{2}a_0(T_s - T_{NI})(\mathbf{Q}:\mathbf{Q}) - \frac{1}{3}b(\mathbf{Q}\cdot\mathbf{Q}):\mathbf{Q} + \frac{1}{4}c(\mathbf{Q}:\mathbf{Q})^2\right) \quad (18)$$

where the right-hand side of eq 18 is evaluated at the bulk nematic phase. Since the isotropic/nematic phase transformation generates heat and increases $T_s$, the interface temperature, the growth kinetics are coupled to heat transfer and it is slowed as $T_s$ increases. This insight into the driving force for spherulite growth in the self-similar regime motivates the inclusion of latent heat effects.

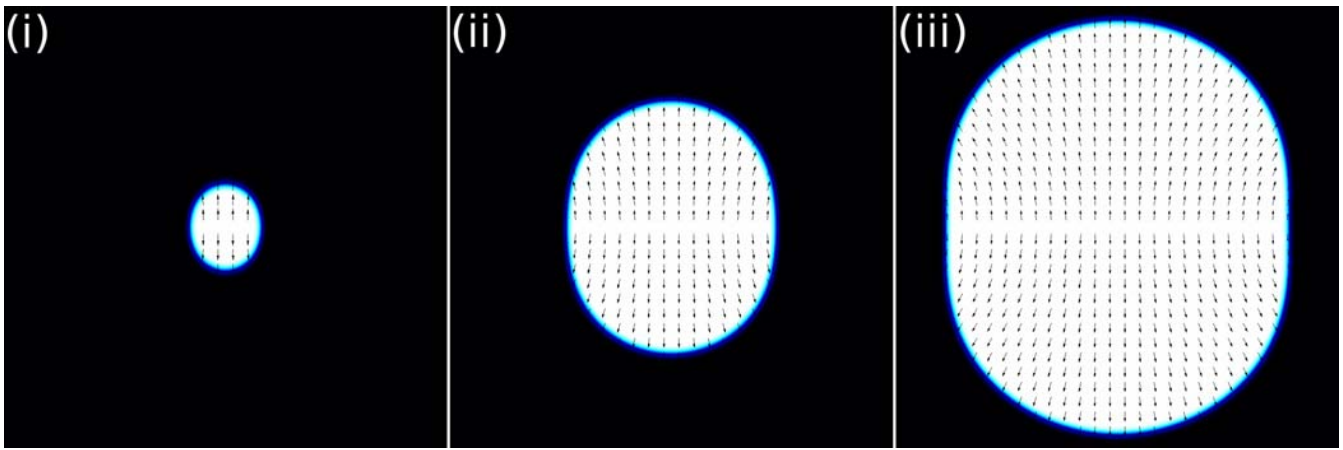

**Figure 5** – Time evolution of a two-dimensional spherulite for a shallow quench (307.3K) in an initially homogeneous configuration of radius 14nm where the surface corresponds to the uniaxial nematic order parameter and the arrows correspond to the nematic director (and should be considered headless); i) initial growth forms an ellipsoidal configuration resulting from interface anisotropy ii-iii) the growth process proceeds in a self-similar fashion; model parameters used in this simulation were $\mu_r$=0.084 N·s/m², $a_0$= 1.4·10⁵ J/m³·K, $b$ = 1.8·10⁷ J/m³, $c$ = 3.6·10⁶ J/m³, $l_1$ = 3.0·10⁻¹² J/m, $l_2$ = 3.1·10⁻¹² J/m, with $T_{NI}$ = 307.2 K[11,12]. Simulation times are i) 0.0586 ms ii) 0.146 ms iii) 0.234 ms and the domain length scale is 3μm. This simulation also captures the characteristics of the non-isothermal initially homogeneous spherulite (NIHS) case, which will be excluded for brevity. Data used for non-isothermal simulations includes $k_{//}$=0.2009 W/m·K, $k_\perp$=0.1364 W/m·K, $C_p$=1800 J/kg·K, $\rho$=1000 kg/m³ from ref 34.

Next, we compare isothermal and non-isothermal growth with initially homogeneous orientation states (IHS and NIHS). Figure 5 shows the isothermal growth of an initially homogeneous spherulite. The isothermal model predicts that, following an initial shape dynamics regime, a self-similar growth regime is observed under a shallow quench. The evolution of the non-isothermal growth of an initially homogeneous spherulite (NIHS) follows the same general behavior as in Figure 5 and is excluded for brevity. The shape dynamic regime and the subsequent final morphology of the spherulite during the self-similar growth regime for this IHS spherulite have been observed in previous work,[9-12] and recent studies have focused on this and similar types of phenomena.[38-40]

Figure 6 shows the power law exponent *n* versus time corresponding to the simulation from Figure 5 simulation (IHS) and the corresponding non-isothermal model results (NIHS). For the diffusionless isothermal model the initial shape dynamic regime and the following self-similar regime are clearly seen with the exponent converging to *n*=1, indicating volume driven growth. For the simulation based on the non-isothermal model, a clear deviation from volumetric driven growth is observed, which will be thoroughly addressed later.

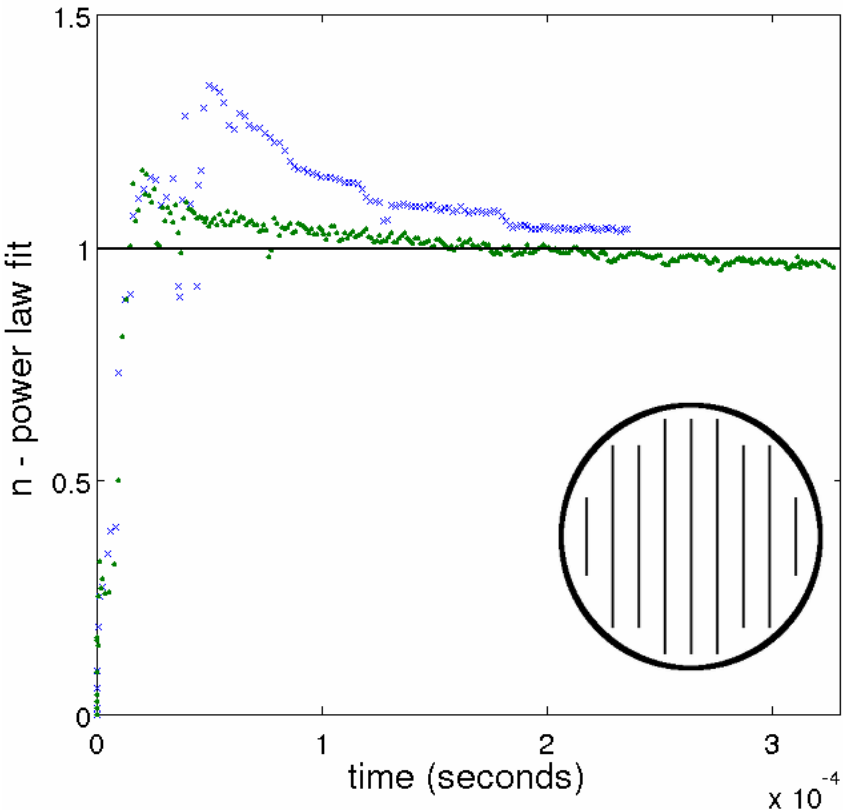

**Figure 6** – Power law exponent versus time for the long axis of: (top curve) isothermal growth with an initially homogeneous spherulite (IHS) and (bottom curve) non-isothermal growth with an initially homogenous spherulite (NIHS)

Next we compare isothermal and non-isothermal growth with initially textured spherulites (ITS and NITS). The growth of a textured spherulite is simpler than that of the homogeneous case in that its morphology is self-similar from the early stages of growth. This type of spherulite morphology typically results from spherulite nucleation due to surface effects of impurities.[41] The estimated critical radius of growth for this spherulite morphology was found to be substantially greater than that of the homogeneous type, most likely due to the requirement of a +1 (or closely coupled pair of +1/2) disclinations at its core. Figure 7 shows simulation results for this spherulite growth process where the initial nucleus configuration splits from a +1 disclination to a pair of +1/2 disclinations during the initial shape dynamics regime. The evolution of the non-isothermal growth of an initially textured spherulite (NITS) follows the same general behavior as in Figure 7 (and thus is not shown), except that the +1 disclination does not split into two +1/2 disclinations. This texture phenomenon resulting from incorporation latent heat effects is interesting, but out of the scope of the growth behavior addressed in this work; refer to refs 35, 36, and 42 for more details on this in an isothermal context. Additionally, the mesh density utilized in these simulations was not adequate to fully resolve disclination defects. The dynamics and topological structure of these defects are dominant only during the shape dynamic growth regime and would not have an effect on the presented results (observed in the self-similar growth regime).

The power law fits for isothermal and non-isothermal growth of textured spherulites are shown in Figure 8. As with the previous spherulite morphology, the trend of the isothermal model is convergence towards volume driven growth, while the non-isothermal model predicts a definitive deviation from that process.

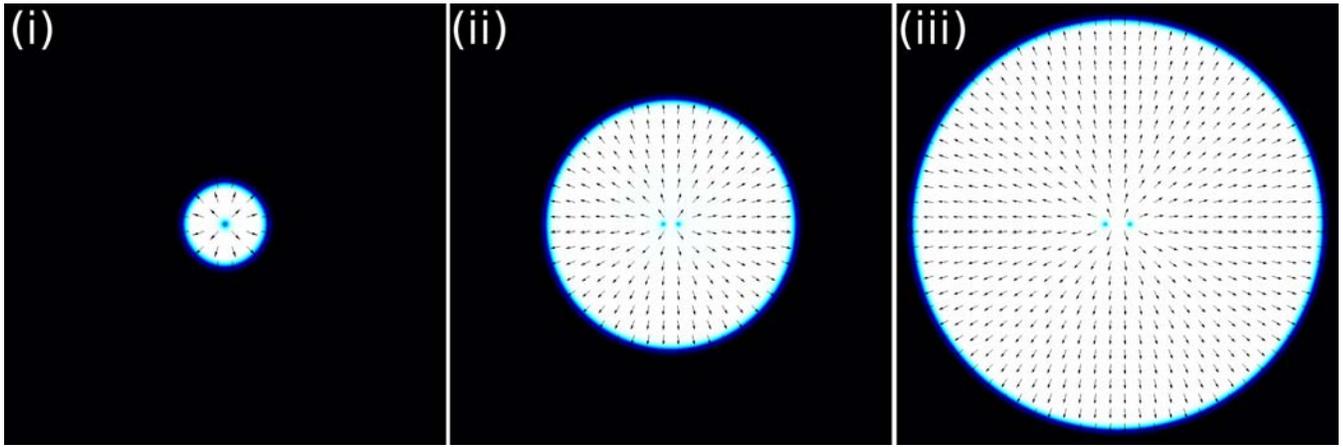

**Figure 7** –Time evolution of a two-dimensional spherulite for a shallow quench (307.3K) in an initially textured spherulite configuration of radius 40nm where the surface corresponds to the uniaxial nematic order parameter and the arrows correspond to the nematic director (and should be considered headless); i) initial growth maintains the initial morphology of a +1 disclination at its core ii) the +1 disclination splits into two +1/2 disclinations ii) the spherulite continues to grow with no interfacial heterogeneity; simulation times are i) 0.0586 ms ii) 0.146 ms iii) 0.234 ms and the domain length scale is 3μm. This simulation also captures the characteristics of the non-isothermal initially textured spherulite (NITS) case except that in this case the +1 disclination does not split into two +1/2 disclinations. This shape dynamic phenomenon is out of the scope of this study (which focuses on growth kinetics) and is excluded for brevity.

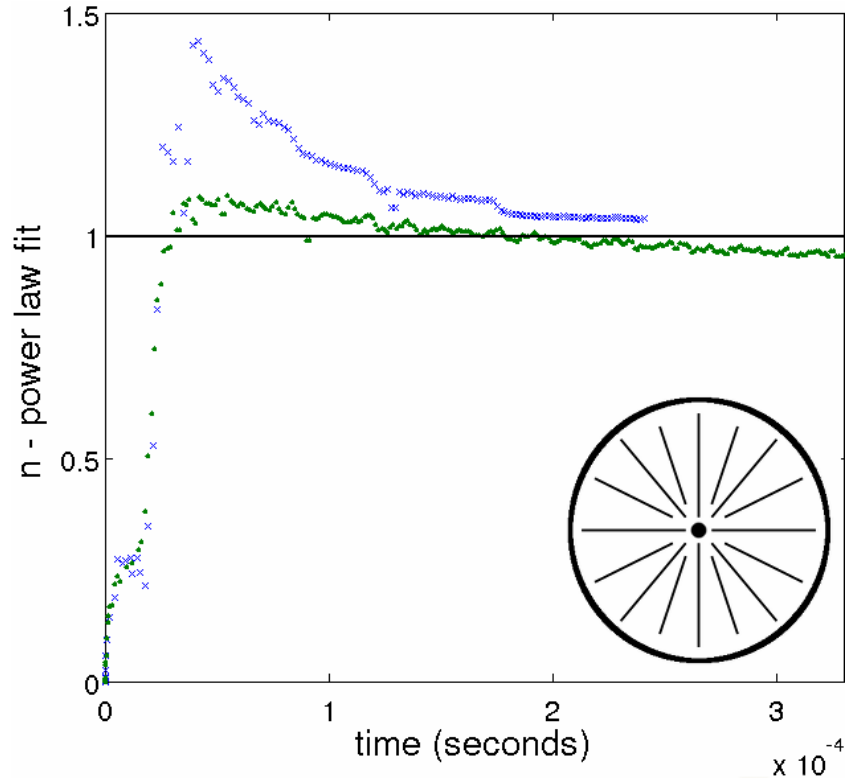

Figure 8 – Power law exponent versus time for: (top curve) isothermal growth with an initially textured spherulite (ITS) and (bottom curve) for non-isothermal growth with an initially textured spherulite (NITS)

Figures 6 and 8 demonstrate that latent heat effects have a non-negligible influence over the self-similar growth regime. These results show that in addition to extending the overall time scale at which the spherulite grows, a deviation from converging to constant interfacial velocity exists as well. As seen in eq 16, this is due to a decrease in the elastic load from the free energy difference decrease.

Latent heat effects modify the spherulite growth predictions in both the initial shape-dynamics regime and the following self-similar growth process. The simulation results of Figures 6 and 8 show that for both cases (NIHS and NITS) the computational domain is not large enough to reach a steady-state thermally limited growth regime. An estimate of the radius at which growth becomes thermal-diffusion limited was found from a simplified analysis of interfacial growth (see Appendix 2) yielding the following expression for the interface temperature:

$$T_s^* = \frac{1}{1 + \frac{2}{R^{*2} Ei(R^*)} \int_{R^*}^{\infty} Ei(r^*) r^* dr^*} \quad (19a)$$

$$r^* = \frac{w}{\alpha} r \,;\, R^* = \frac{w}{\alpha} R \,;\, T_s^* = \frac{T_s - T_\infty}{G/\rho C p} \quad (19b)$$

where $\alpha = k/\rho \cdot C_p$ is the thermal diffusivity, $T_\infty$ is the temperature at infinite radius, and $Ei$ is the exponential integral.

For an initial temperature of $T$=307.30 K, we can calculate the heat generatd by the phase transition from eq 12, considering that the equilibrium order parameter in the nematic phase is approximately 0.20, and find $G$=5.8·10$^5$ J/m$^3$. In order to estimate the radius at which growth becomes heat diffusion-limited, we can assume that the spherulite grows at constant velocity until its temperature reaches the phase coexistence temperature. In actuality, the transition takes place gradually over a broad radius range where the velocity changes smoothly, the beginning of which is seen from simulation results in Figures 6 and 8.

For the 5CB parameters used (see Figure 5), the coexistence temperature is 307.44 K. Thus from eq 19b, $T_s^* = 0.45$ and from eq 19a, the dimensionless radius is $R^*=1.6$. Using the velocity corresponding to the stationary growth at 307.3 K, which is obtained from simulation to be approximately 0.006 m/s, the estimated thermal diffusion limited growth radius is $R = 19$ μm, which is significantly larger than our simulation domain (R=1.5μm). In order to observe this transition of the growth process while accommodating computational constraints, an artificial reduction of the value of the thermal conductivity can be used. This material property change (i.e. lower **K** from eq 11) effectively decreases the thermal-diffusion limited growth radius. Based upon the previously derived approximation, for the artificial 5CB system with **K'=K/100**, the spherulite radius must be on the order of $R = 0.19$ μm before thermally limited growth is observed.

Figure 9a shows the power law exponent from simulations of the homogeneous cases (from Figure 6) and, in addition, from simulations with the artificial 5CB parameters. Figure 9b shows the power law exponent from simulations of the textured cases (from Figure 8) and, in addition, from simulation with the artificial 5CB parameters. As expected, latent heat effects are magnified as a result of the reduced thermal diffusivity, and thermally limited growth appears at smaller spherulite radii.

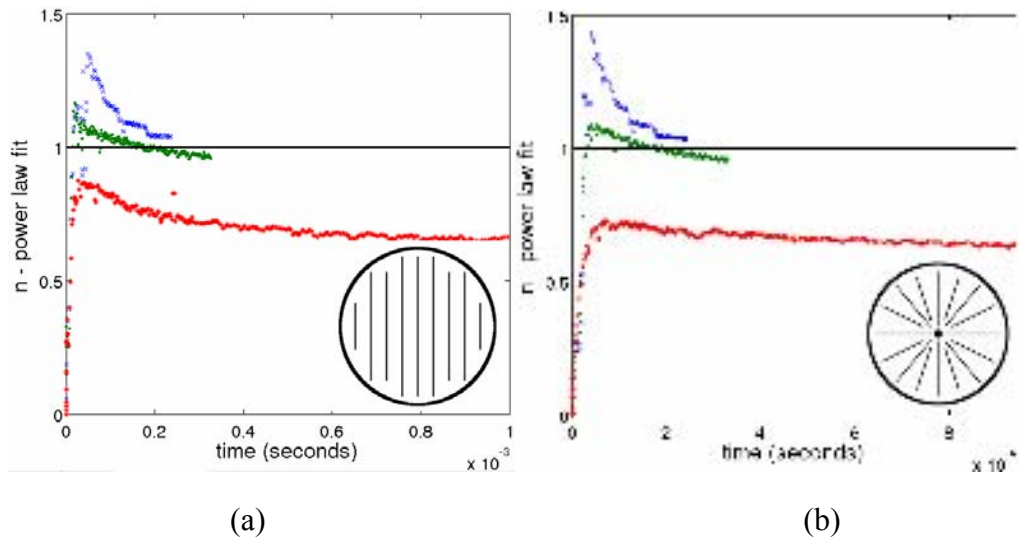

(a)  (b)

**Figure 9**. Power law exponent *n* versus time for a) homogeneous spherulite growth configurations for modified 5CB (bottom), actual 5CB parameters with non-isothermal growth (middle), and actual 5CB parameters with isothermal growth (top) b) textured spherulite growth configurations for modified 5CB (bottom), actual 5CB parameters with non-isothermal growth (middle), and actual 5CB parameters with isothermal growth (top).

Confirmation of the cause for thermally limited growth can be seen from the temperature profile of the interface between the nematic and isotropic phases for the artificial 5CB parameters from Figure 9. Figure 10 shows the temperature profile from the center of the spherulite towards the far field occupied by the isotropic phase, at different times in the post-shape dynamic growth regime. The dark parabolic line is the interface temperature, and the profile lines correspond to the temperature profile across the interface. As time increases the temperature within the spherulite and at the interface increases. The diffusive process is small enough in magnitude such that the local interfacial temperature approaches the bulk transition temperature. Approaching this temperature the free energy difference between the isotropic and nematic phases approaches zero and interfacial growth stagnates. Thus, as the interface temperature approaches the bulk transition, a steady state is reached as the heat removed from the interface diffuses at the same rate at which it is generated.

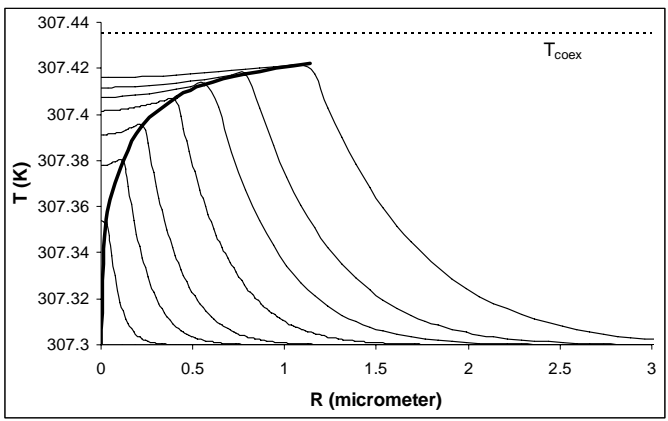

**Figure 10** – Temperature profile at different times of interface growth with the maximum temperature at each step outlined. The bulk transition temperature for the parameters used (307.44K) is outlined and the trend of the interface temperature to approach it is clearly shown.

### CONCLUSIONS

In this work simulation of the growth of a two-dimensional nematic spherulite in an isotropic matrix was presented. A thermal energy balance was derived and used to incorporate latent heat effects of the phase ordering transition on interfacial growth. The main contributions of this work are:

- An anisotropic thermal energy balance, eq 9, was derived (see Appendix 1) that incorporates latent heat effects through coupling with the nematic free energy density, eq 7.

- Simulation results for both homogeneous and textured (see Figure 4) nematic spherulites were performed with and without the thermal energy balance, showing that inclusion of the energy balance is required to predict a self-similar growth regime with a growth exponent $n$<1 at shallow quench depths, as observed in experimentally.[17-22]

- Thermal effects were found to increase the timescales across the full range of the spherulite growth process, including the initial shape dynamic regime (see Figures 6 and 8).

While computational limitations do need to be addressed in the future, the application of the presented model to many complex physical systems is possible. These systems include those aligned with industrial applications, such as confined nematics, and more theoretical applications such as the study of thermal instabilities in nematic interface growth. This work has set a basis for these types of future applications and for the prediction of phenomena resulting from the interplay of phase ordering and thermal effects.

### ACKNOWLEDGMENT

This research was supported in part by the Natural Science and Engineering Research Council of Canada (NMA and ADR) and the National Research Council of Argentina (ERS).

# APPENDIX 1

The energy balance for a differential volume, without flow, is:

$$\frac{du}{dt} = \frac{dW}{dt} - \nabla \cdot \mathbf{q} \tag{A1.1}$$

where $u$ is the internal energy density, $\mathbf{q}$ is the heat flux, and $dW/dt$ is the rate of mechanical work on the differential control volume. According to eq. 4.46 of ref 43, the rate of mechanical work on the system can be written as the sum of elastic energy stored and the rate of dissipation. The rate of elastic energy storage is the change in nematic energy due to changes in the order parameter, so it is $df_n/dt$ taken at constant temperature. The rate of dissipation in the absence of flow is equal to $\mathbf{H}{:}d\mathbf{Q}/dt$,[44] where the molecular field $\mathbf{H}$ is the negative of the variational derivative of $\mathbf{Q}$. With no flow, the Landau - de Gennes model gives:

$$\mathbf{H} = \mu \frac{\partial \mathbf{Q}}{\partial t} = -\left.\frac{\partial f_n}{\partial \mathbf{Q}}\right|_{T,\nabla\mathbf{Q}} + \nabla \cdot \left.\frac{\partial f_n}{\partial \nabla\mathbf{Q}}\right|_{T,\mathbf{Q}} \tag{A1.2}$$

The rate of mechanical work is thus:

$$\begin{aligned}\frac{dW}{dt} &= \left.\frac{\partial f_n}{\partial \mathbf{Q}}\right|_{T,\nabla\mathbf{Q}} : \frac{\partial \mathbf{Q}}{\partial t} + \left.\frac{\partial f_n}{\partial \nabla\mathbf{Q}}\right|_{T,\mathbf{Q}} : \frac{\partial \nabla\mathbf{Q}}{\partial t} + \left(-\left.\frac{\partial f_n}{\partial \mathbf{Q}}\right|_{T,\nabla\mathbf{Q}} + \nabla \cdot \left.\frac{\partial f_n}{\partial \nabla\mathbf{Q}}\right|_{T,\mathbf{Q}}\right) : \frac{\partial \mathbf{Q}}{\partial t} \\ &= \left.\frac{\partial f_n}{\partial \nabla\mathbf{Q}}\right|_{T,\mathbf{Q}} : \frac{\partial \nabla\mathbf{Q}}{\partial t} + \nabla \cdot \left.\frac{\partial f_n}{\partial \nabla\mathbf{Q}}\right|_{T,\mathbf{Q}} : \frac{\partial \mathbf{Q}}{\partial t}\end{aligned} \tag{A1.3}$$

The left-hand side of eq A1.1 can be written in terms of Helmholtz free energy:

$$\frac{\partial u}{\partial t} = \frac{\partial}{\partial t}\left(f - T\left.\frac{\partial f}{\partial T}\right|_{\mathbf{Q},\nabla\mathbf{Q}}\right) = \frac{\partial f}{\partial t} - \left.\frac{\partial f}{\partial T}\right|_{\mathbf{Q},\nabla\mathbf{Q}} \frac{\partial T}{\partial t} - T\frac{\partial}{\partial t}\left.\frac{\partial f}{\partial T}\right|_{\mathbf{Q},\nabla\mathbf{Q}} \tag{A1.4}$$

Expanding the time derivatives:

$$\frac{\partial f}{\partial t} = \left.\frac{\partial f}{\partial T}\right|_{\mathbf{Q},\nabla\mathbf{Q}} \frac{\partial T}{\partial t} + \left.\frac{\partial f}{\partial \mathbf{Q}}\right|_{T,\nabla\mathbf{Q}} \frac{\partial \mathbf{Q}}{\partial t} + \left.\frac{\partial f}{\partial \nabla\mathbf{Q}}\right|_{T,\mathbf{Q}} \frac{\partial \nabla\mathbf{Q}}{\partial t} \tag{A1.5a}$$

$$-T\frac{\partial}{\partial t}\left.\frac{\partial f}{\partial T}\right|_{\mathbf{Q},\nabla\mathbf{Q}} = -T\left(\left.\frac{\partial^2 f}{\partial T^2}\right|_{\mathbf{Q},\nabla\mathbf{Q}}\right)\frac{\partial T}{\partial t} - T\frac{\partial}{\partial \mathbf{Q}}\left(\left.\frac{\partial f}{\partial T}\right|_{\mathbf{Q},\nabla\mathbf{Q}}\right)\frac{\partial \mathbf{Q}}{\partial t} \tag{A1.5b}$$

In equation A1.5b, it has been taken into account that:

$$\frac{\partial}{\partial \nabla \mathbf{Q}}\left(\left.\frac{\partial f}{\partial T}\right|_{\mathbf{Q},\nabla \mathbf{Q}}\right) = 0 \qquad (A1.6)$$

due to the assumption that the elastic constants are independent of temperature. Inserting eqs A1.5a and A1.5b in A1.4, and taking into account that the specific heat (per unit volume) is:

$$C_p = -T\left(\left.\frac{\partial^2 f}{\partial T^2}\right|_{\mathbf{Q},\nabla \mathbf{Q}}\right) \qquad (A1.7)$$

The total change of internal energy is:

$$\frac{\partial u}{\partial t} = C_p \frac{\partial T}{\partial t} + \left.\frac{\partial f}{\partial \mathbf{Q}}\right|_{T,\nabla \mathbf{Q}} \frac{\partial \mathbf{Q}}{\partial t} + \left.\frac{\partial f}{\partial \nabla \mathbf{Q}}\right|_{\mathbf{Q},T} \frac{\partial \nabla \mathbf{Q}}{\partial t} - T\frac{\partial}{\partial \mathbf{Q}}\left(\left.\frac{\partial f}{\partial T}\right|_{\mathbf{Q},\nabla \mathbf{Q}}\right)\frac{\partial \mathbf{Q}}{\partial t} \qquad (A1.8)$$

Combining A1.8, A1.3 and A1.1, the final expression for the energy balance is:

$$C_p \frac{\partial T}{\partial t} = \left(-\left.\frac{\partial f}{\partial \mathbf{Q}}\right|_{T,\nabla \mathbf{Q}} + \nabla \cdot \frac{\partial f}{\partial \nabla \mathbf{Q}}\right):\frac{\partial \mathbf{Q}}{\partial t} + T\frac{\partial}{\partial \mathbf{Q}}\left(\left.\frac{\partial f}{\partial T}\right|_{\mathbf{Q},\nabla \mathbf{Q}}\right):\frac{\partial \mathbf{Q}}{\partial t} - \nabla \cdot \mathbf{q} \qquad (A1.9)$$

## APPENDIX 2

The purpose of this Appendix is to derive eq 19a. An approximation of the radius at which growth becomes thermal-diffusion limited is obtained from a simplified analysis of the interface growth process. Assuming that the spherulite has a sharp interface, an integral energy balance for the spherulite and the surrounding medium can be derived, noting that the heat generated by the phase transition increases the temperature of the nematic spherulite and the isotropic medium. Considering a uniform temperature in the nematic spherulite, we get:

$$V\rho C_p (T_s - T_0) + \Delta H_{iso} = GV \tag{A2.1}$$

where $T_s$ is the spherulite temperature, $T_0$ is the initial temperature, $\Delta H_{iso}$ is the enthalpy change in the isotropic phase (due to the phase transition), $G$ the heat generated by the phase transition, and $V$ is the spherulite volume. As the temperature in the isotropic phase is not uniform, the total enthalpy increase in the isotropic phase is: $\int_R^\infty \rho C_p (T(r) - T_\infty) 2\pi r dr$, where $T_\infty$ is the temperature at infinite radius. After rearranging and noting that $T_\infty = T_0$:

$$(T_s - T_\infty) + \int_R^\infty (T(r) - T_\infty) \frac{2r}{R^2} dr = \frac{G}{\rho C_p} \tag{A2.2}$$

where $R$ is the spherulite radius. The solution of this equation requires specification of the temperature distribution $T(r)$ in the isotropic phase. The differential energy balance for the isotropic phase in two-dimensions, assuming radial symmetry, is:

$$\frac{\partial T}{dt} = \alpha \frac{1}{r} \frac{\partial}{\partial r} \left( r \frac{\partial T}{dr} \right) \tag{A2.3}$$

Changing coordinates to a frame moving with the interface, $r' \longrightarrow r - R$:

$$\frac{\partial T}{dt} - w \frac{\partial T}{dr'} = \alpha \frac{1}{r' + R} \frac{\partial}{\partial r'} \left( (r' + R) \frac{\partial T}{dr'} \right) \tag{A2.4}$$

Note that $R = R(t) = \int w \, dt$, where $w$ is the velocity of the moving interface and it can be non-constant.

This equation allows a pseudo-stationary solution where $\frac{\partial T}{\partial t} = 0$:

$$T - T_\infty = -C \int_\infty^{r'+R} \frac{e^{-\frac{v}{\alpha}a}}{a} da = C \, Ei\left(\frac{v}{\alpha}(r'+R)\right) \tag{A2.5}$$

where $C$ is an integration constant and $a$ is a dummy integration variable. $Ei$ is the exponential integral, and can be approximated with the following analytical expression:

$$Ei(x) = -0.57721 - \ln(x) + \sum_1^\infty \frac{x^n}{n\,n!} \tag{A2.6}$$

Applying the boundary condition at the interface ($T = T_s$ at $r' = 0$) yields:

$$T(r) - T_\infty = \frac{T_s - T_\infty}{Ei\left(\frac{w}{\alpha}R\right)} Ei\left(\frac{w}{\alpha}(r'+R)\right) = \frac{T_s - T_\infty}{Ei\left(\frac{w}{\alpha}R\right)} Ei\left(\frac{w}{\alpha}r\right) \tag{A2.7}$$

Assuming a pseudo-steady state temperature distribution in the isotropic phase, eq A2.2 becomes:

$$(T_s - T_0) + \int_R^\infty \frac{T_s - T_\infty}{Ei\left(\frac{w}{\alpha}R\right)} Ei\left(\frac{w}{\alpha}r\right) \frac{2r}{R^2} dr = \frac{G}{\rho C_p} \tag{A2.8}$$

Utilizing non-dimensional variables $r^* = \frac{w}{\alpha}r$, $R^* = \frac{w}{\alpha}R$ and $T_s^* = \frac{(T_s - T_\infty)\rho C_p}{G}$, the dimensionless form of eq A2.8 equation is

$$T_s^* = \frac{1}{1 + \frac{2}{R^{*2} Ei(R^*)} \int_{R^*}^\infty Ei(r^*) r^* dr^*} \tag{A2.9}$$

Table of Contents Graphic

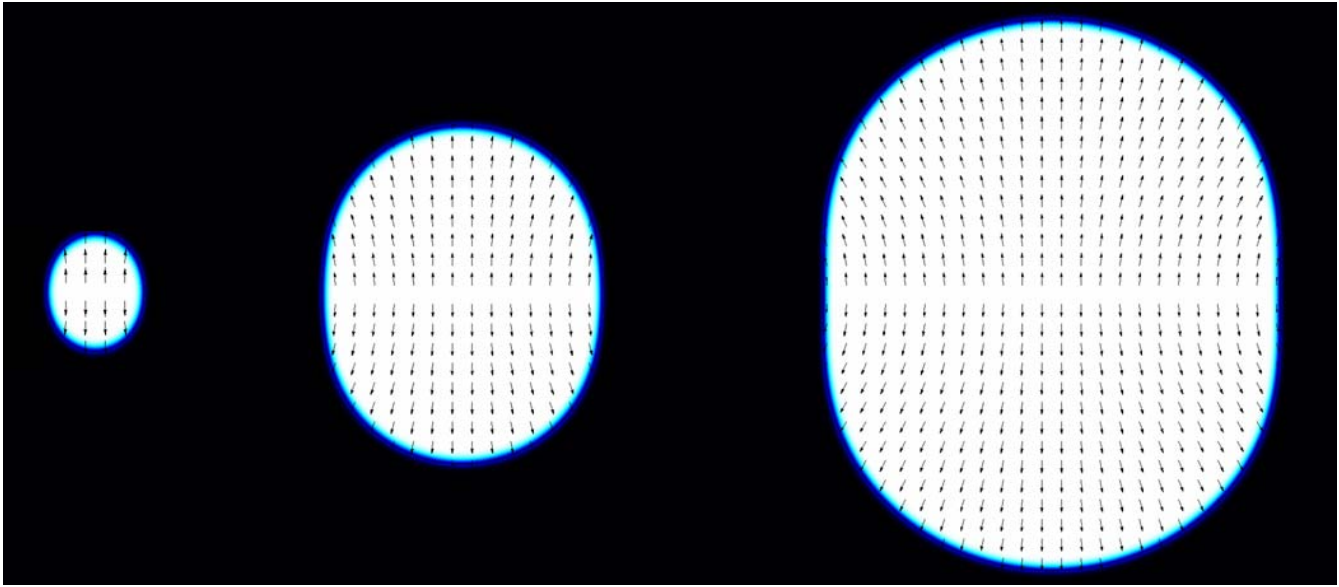